\documentclass[%
 aip,
 amsmath,amssymb,
 reprint,%
]{revtex4-1}

\usepackage{graphicx}
\usepackage{dcolumn}
\usepackage{bm}

\usepackage[utf8]{inputenc}
\usepackage[T1]{fontenc}
\usepackage{mathptmx}
\usepackage{etoolbox}
\usepackage{xcolor}

\usepackage{comment}

\makeatletter
\def\@email#1#2{%
 \endgroup
 \patchcmd{\titleblock@produce}
  {\frontmatter@RRAPformat}
  {\frontmatter@RRAPformat{\produce@RRAP{*#1\href{mailto:#2}{#2}}}\frontmatter@RRAPformat}
  {}{}
}%
\makeatother
\begin{document}

\preprint{AIP/123-QED}

\title[Polar vs nematic unjamming]{Distinct impacts of polar and nematic self-propulsion on active unjamming}
\author{Varun Venkatesh}
\altaffiliation{Both authors contributed equally to this manuscript}
\affiliation{Niels Bohr Institute, University of Copenhagen, Blegdamsvej 17, 2100, Copenhagen, Denmark.}

\author{Chandana Mondal}
\altaffiliation{Both authors contributed equally to this manuscript}
\affiliation{%
UGC-DAE Consortium for Scientific Research, University Campus, Khandwa Road, Indore 452017, India}
%

\author{Amin Doostmohammadi*}
 \email{doostmohammadi@nbi.ku.dk}
\affiliation{Niels Bohr Institute, University of Copenhagen, Blegdamsvej 17, 2100, Copenhagen, Denmark.}


\begin{abstract}
 Though jamming transitions are long studied in condensed matter physics and granular systems, much less is known about active jamming (or unjamming), which commonly takes place in living materials. In this paper, we explore, by molecular dynamic simulations, the jamming-unjamming transition in a dense system of active semi-flexible filaments. In particular we characterise the distinct impact of polar versus nematic driving for different filament rigidity and at varying density. Our results show that high densities of dynamic active filaments can be achieved by only changing the nature of the active force, nematic or polar. Interestingly, while polar driving is more effective at unjamming the system at high densities below confluency, we find that at even higher densities nematic driving enhances unjamming compared to its polar counterpart. The effect of varying the rigidity of filaments is also significantly different in the two cases: while for nematic driving lowering bending rigidity unjams the system, we find an intriguing re-entrant jamming-unjamming-jamming transition for polar driving as the filament rigidity is lowered. While the first transition (unjamming) is driven by softening due to reduced rigidity, the second transition (jamming) is a cooperative effect of ordering and coincides with the emergence of nematic order in the system. Together, through a generic model of self-propelled flexible filaments, our results demonstrate how tuning the nature of self-propulsion and flexibility can be employed by active materials to achieve high densities without getting jammed.
\end{abstract}

\maketitle

\section{Introduction}
Spontaneous collective motion and self-organisation of living materials play a crucial role in diverse biological phenomena from embryo-genesis ~\cite{friedl_collective_2009, pawlizak_testing_2015,mongera_fluid--solid_2018}, tissue repair ~\cite{petridou_fluidization-mediated_2019, petitjean_velocity_2010}, to bacterial invasion and biofilm formation \cite{hartmann_emergence_2019, kempf_active_2019}, subcellular organization of actin and microtubule filaments \cite{jalal_actin_2019, sanchez_spontaneous_2012}, and in collective sperm motility in reproductive tracts\cite{schoeller_collective_2020}. At high densities of living materials these collective motions often lead to jamming and unjamming of the system, where the material transitions between a solid-like jammed state to a fluid-like moving state. Notable biological examples include the extensively studied epithelial cells ~\cite{park_unjamming_2015, atia_geometric_2018, angelini_glass-like_2011, mitchel_primary_2020, malinverno_endocytic_2017}, fibroblast cells~\cite{duclos_perfect_2014}, bacterial biofilm formation ~\cite{delarue_self-driven_2016, worlitzer_biophysics_2021}, and sperm cells~\cite{gervasi_molecular_2017}. While some biological processes, e.g. embryogenesis ~\cite{blauth_jamming_2021}, and wound healing ~\cite{nnetu_impact_2012} are driven by solidification of the living matter, which can be linked to a jamming transition, some other, such as cancer progression occur due to fluidization (unjamming) of the matter~\cite{oswald_jamming_2017, ilina_cellcell_2020}. A rigorously developed framework is therefore needed to establish a unifying understanding of the physics of jamming-unjamming transition in active matter and the generic features that control it.

Jamming and unjamming has been studied in great detail in granular systems for both passive \cite{zhang_jamming_2005} \cite{majmudar_jamming_2007} and active particles, where jamming is seen to occur at high packing fractions \cite{mandal_extreme_2020} and low self-propulsion \cite{henkes_active_2011}.  While features such as density and particle aspect ratio impact both passive \cite{marschall_compression-driven_2018, marschall_shear-driven_2019} and active jamming \cite{wensink_emergent_2012}, it is important to understand that the jamming picture is much simpler in case of passive systems. The main parameter controlling passive jamming is the number density or the packing fraction. As the density increases, the particles get more and more constrained by their neighbours. The higher the density, the larger the energy barrier associated with structural rearrangements of the particles. Above the critical density, $\rho_C$, a geometrical transition, called the jamming transition, occurs as the particle motion becomes caged and the system shows solid-like behaviour. In active matter, the jamming phase diagram is more complex due to the presence of additional controlling factors. Like inert systems, density can also lead to jamming transitions in active systems. However, other factors such as self-propulsion, active interaction between particles \cite{garcia_physics_2015} and active cell shape changes \cite{park_unjamming_2015} can sometimes impact the jamming transition more than density. In this vein, we note that technically, jamming is a geometric transition and occurs in the absence of dynamics, while active or thermal driving result in glassy dynamics. However, jamming-unjamming has been used to refer to the transition from a solid-like to fluid-like state even in the presence of activity~\cite{park_unjamming_2015}. A striking feature of active systems is their ability to continuously produce work, through their activity, which helps them maintain their flow even at very high packing fractions. However, it is not clear how different forms of activity impact jamming-unjamming transitions in dense systems and whether switching between different activity modes can affect the jamming density in active materials.

Common to many active systems is the ability of their self-propelled constituents to frequently change their direction of motion. A well-known example of this is the run-and-tumble motion~\cite{turner_real-time_2000, tailleur_statistical_2008}. This active directional change of motion is an integral part of active particle motion driven by external stimuli such as food, chemical, or optical stimuli. While some bactria, like swimming \textit{Escherichia coli}, randomly change their direction of motion by some finite angle, other systems show a complete $180^o$ change in direction. This is ubiquitously seen in various bacterial swarms including in the gliding motility of \textit{M. xanthus}~\cite{starrus_pattern-formation_2012}, in gliding filamentous Cyanobacteria ~\cite{tamulonis_modeling_2011}, invasive colonies of \textit{Pseudomonas ariginosa}~\cite{meacock_bacteria_2021}, swimming soil bacteria \textit{Pseudomonas putida}~\cite{theves_bacterial_2013}, as well as in turbulent swarming of sperm cells \cite{creppy_turbulence_2015}, and elongated neural progenitor stem cells~\cite{kawaguchi_topological_2017}. The reversal time is of biological relevance, for example in M. xanthus, where the most efficient spread of extracellular protein is observed for bacteria that reverse directions neither too fast nor too slow \cite{amiri_reversals_2017}.

Importantly, these directional changes can affect the type of forces that active agents exert on their surrounding. Previous studies have considered collective patterns of motion in the two limits of extremely slow or extremely fast reversals, where the self-propulsion forces correspond to polar~\cite{duman_collective_2018} and nematic (apolar)~\cite{joshi_interplay_2018} driving, respectively. The former corresponds to persistent directional self-propulsion, while the latter is the result of fast and highly frequent motion reversals. In dilute systems, polar filaments exhibit spiralling and clustering that is activity and flexibility dependent~\cite{duman_collective_2018}, while nematic filaments display a reduction in clustering and spiral formation as shown by a recent study comparing polar and nematic filaments at low density~\cite{abbaspour_effects_2021}.
Polar filaments, at high densities, show a jamming transition that is activity dependent, while for densely packed nematic filaments~\cite{duman_collective_2018}, the flexibility itself has been shown to be dependent on the activity, both in simulation~\cite{joshi_interplay_2018} and experiments~\cite{kumar_tunable_2018}. Nevertheless, the distinct impacts of polar and nematic driving on jamming-unjamming transition of active flexible filaments at high densities is yet to be explored. 

Here, we characterise the impact of active and passive micro-structural features on the unjamming of dense active filaments using a semi-flexible self-propelled worm-like chain model. Through out the study, we contrast the transition for polar filament with that of nematic filaments for different filament flexibility and variable packing fraction, consistently showing that the jamming- unjamming transition is markedly different in the two cases. We are able to tune the polarity of the filaments by controlling reversal time for the active driving force. 
We find a remarkable difference between polar and nematic filaments when we study the jamming/unjamming transition in the rigidity-activity plane. While for nematic filaments the activity required for unjamming monotonically decrease with decreasing rigidity, we find a re-entrant jamming transition at lower rigidity for polar filaments. 
Interestingly, we show that the efficacy of polar or nematic driving in unjamming an active jammed system is density-dependent: at densities close to typical packing fractions for passive systems, nematic filaments require more activity to unjam compared to polar filaments; however, this trend changes significantly in the very high density regime where we find that the activity required for unjamming in polar filaments is far greater than that for nematic filaments. 

The rest of this paper is organised as follows. In section 2 we introduce the model describing different filament attributes which set the parameter space for our study and give details of the molecular dynamic simulations. We present our main findings from our simulations, where we characterise the effect of different filament attributes to the jamming/unjamming transition of the system in section 3.  Finally, we briefly discuss some of the biological implications of our findings.
\section{Numerical model and methods}
\begin{figure}[b]
	\includegraphics[scale=0.5]{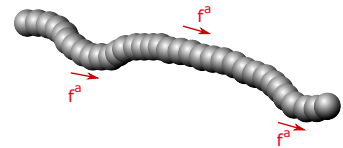}
	\caption{Filament model: beads are connected via stiff springs. The active force acts tangentially along all bonds as shown by the red arrows. To continuously crossover from polar to nematic driving, the head and tail of each filament are switched at stochastic, Poisson distribution intervals with mean $\tau_n$ so that the motion reversal occurs at the specified time scale $\tau_n$.}
	\label{filament}
\end{figure}
\begin{figure*}[ht]
    \centering
    \includegraphics[scale=0.6]{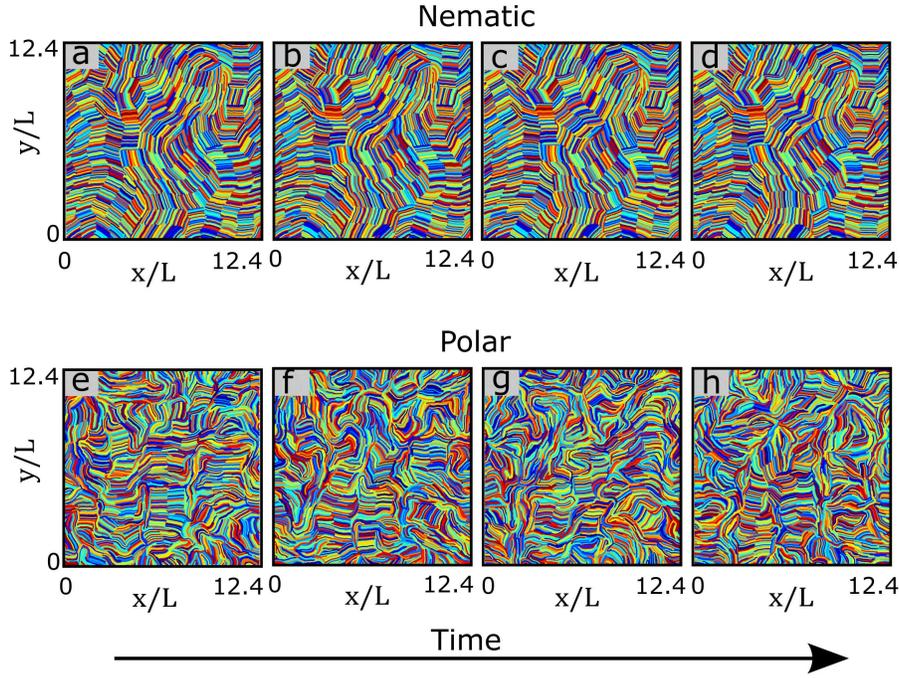}
    \caption{Snapshots of nematic (a-d) and polar (e-h) filaments showing the jammed and unjammed state at four times: $t = 173.0\tau$ (a, e), $t=207.6\tau$ (b, f), $t=242.2\tau$ (c, g), and $t = 276.8\tau$ (d, h). The activity is set as $Pe = 1.7$ (see also Supplementary Movies~\cite{smovie1}).}
    \label{fig:1}
\end{figure*}
%
\subsection{\label{sec:model}Model}
We study a system consisting of self-propelled worm-like semi-flexible filaments in 2-dimensions. Each filament is modelled as bead-spring polymer consisting of $N+1$ beads of diameter $\sigma$, held together by $N$ stiff harmonic bonds and bending potentials. Fig.~\ref{filament} shows a single filament consisting of 35 beads. Activity is implemented as a force of magnitude $f^a$ acting on each filament bead along the local tangent as indicated by the  red arrows in the figure. $L$ is the length of the filament. The Langevin equations governing the system dynamics were integrated using LAMMPS \cite{thompson_lammps_2022} with in-house modifications. The model combines features of self-propelled worm-like chain model developed recently \cite{eisele-holder_self-propelled_2015} and the reversing rods approach to rigid active nematic rods \cite{joshi_interplay_2018}. The equation of motion for bead $i$ is given by the Langevin equation
\begin{equation}
m{\bf \ddot{r_i}}=-\gamma {\bf \dot{r}}_i - {\bf \nabla}_i U + {\bf f}^{k_BT}_i + {\bf f}_i^a
\label{eqmot}
\end{equation}
 where $m$ is the mass of each bead, ${\bf {r}_i}$ is the position, $\gamma$ denotes the friction coefficient, "dots" denote derivatives with respect to time. ${\bf f}^{k_BT}_i$  denotes the delta-correlated thermal noise force acting on particle $i$ with zero mean and variance $2k_BT\gamma/ \Delta t$, to fulfil the fluctuation–dissipation theorem. $T$ is the temperature. ${\bf f}_i^a$ is the active force on bead $i$ which is a propulsive force on every bead directed along the local tangent. To render the activity nematic, the head and tail of each filament are switched at stochastic, Poisson distributed intervals with mean $\tau_n$ so that the force direction on each bead rotates by $\pi$. In particular, the active force has the form,
\begin{equation}
{\bf f}_i^a = \eta(t) f^a \frac{{\bf r}_{i,i+1}}{|{\bf r}_{i,i+1}|}
\label{eqnem}
\end{equation}
where $f^a$ parameterizes the activity and the bond vector between bead $i$ and $i+1$ is defined as ${\bf r}_{i,i+1} = {\bf r}_{i+1} - {\bf r}_i$. $\eta(t)$ is a stochastic variable associated with each filament  (same for all beads in a filament) that changes its values between $1$ and -$1$ on Poisson distributed intervals with mean $\tau_n$, so that $\tau_n$ controls the reversal frequency. $\gamma {\bf \dot{r}}_i$ is the viscous drag force. $U$ is the interaction potential given by,
\begin{equation}
U = U_{bond} + U_{bend} + U_{EV}
\label{U}
\end{equation}
It consists of bond, bending and excluded-volume contributions.
The harmonic bond potential,
\begin{equation}
U_{bond} = \frac{k_H}{2}\sum_j\sum_{i=1}^{N}\left[|{\bf r}_{i,i+1}^j|-b\right]^2
\label{Ubond}
\end{equation}
acts on the neighboring beads of each filament separately with $j$ and $i$ denoting the filament index and the bead index of a filament, respectively.  $k_H$ is the harmonic spring constant for the bond potential and $b$ is the equilibrium bond length, which is the same for all the bonds. We take $k_H$ to be sufficiently large to keep the bond-lengths essentially fixed at $b$ and choose $b=\sigma/2$ to make the filaments smooth. Semi-flexibility is introduced via the bending potential,
\begin{equation}
U_{bend} = \frac{k_A}{4b^2}\sum_j\sum_{i=1}^{N-1}\left[{\bf r}_{i,i+1}^j-{\bf r}_{i+1,i+2}^j\right]^2
\label{Ubend}
\end{equation}
where $k_A$ is the bending rigidity. The excluded-volume interaction acts between the beads of a single filament and between the beads of different filaments alike to render the filaments impenetrable. It is modelled by the Lenard-Jones (LJ) potential,
\begin{equation}
U_{EV} = \sum_i\sum_{j>i}u_{EV}({\bf r}_{i,j})
\label{UEV}
\end{equation}
where,
\begin{equation}
u_{EV}(r) = \left \{  
\begin{array}{lr}
4\epsilon \left [ \left ( \frac{\sigma}{r} \right )^{12}- \left ( 
\frac{\sigma}{r} \right )^{6}\right ]+\epsilon,\quad r< 
r_c \\ 
0, \quad r\geq r_c
\end{array}
 \right.
\label{uEV}
\end{equation}
where, ${\bf r}_{i,j}={\bf r}_i -{\bf r}_j$ is the vector connecting the positions of the beads $i$ and $j$ (which may belong to the same or different filaments). $\epsilon$ is the characteristic volume-exclusion energy and $r_c$ is the cutoff. The potential is truncated and shifted to zero at $r_c$. Unless otherwise stated $r_c=2^{1/6}\sigma$ for our simulations which makes the interaction potential completely repulsive. 
%
\subsection{Parameters}
In our simulations the dynamics of a single filament is mainly controlled by a dimensionless number, the P\'eclet number $Pe = \frac{f^aL}{k_BT}$, defined as the ratio between advective and diffusive transport, thereby providing a measure for the strength of self-propulsion. Here $L=Nb$, the length of a single filament which is shown in Fig.~\ref{filament}. Another important parameter in our model is the bending rigidity $k_A$. Accordingly, one can define another dimensionless number $\xi_P/L=\frac{k_A}{k_BT}$, where $\xi_P$ is the thermal persistence length. A higher value of $\xi_P/L$ means a stiffer filament.

The results are presented in dimensionless units, with length measured in units of $L$, energies in units of the thermal energy $k_BT$, and time in units of  $\tau=\gamma L^2/k_BT$. Equations of motion are integrated with the Verlet algorithm using LAMMPS. We used mass $m=1$ for all beads. Friction coefficient $\gamma$ is chosen such that the dynamics is close to over-damped dynamics. 

Unless otherwise stated, we use $k_H = 5000k_BT/\sigma^2$, and $\epsilon = k_BT$. We consider a square box of lengths $L_x=L_y=211.25 \sigma$ with periodic boundaries along both $x$ and $y$ directions. We generate the initial configuration by first placing the filaments in completely extended configurations (all bonds parallel and all bond lengths set to $b$) at the desired packing fraction which is defined as $\phi = N_bN_f\sigma b/L_xL_y$, $N_f =2100$ being the number of filaments and $N_b=34$ being the number of bonds. The aspect ratio $a=L/\sigma$ is fixed at $a=17$ in our simulations. The parameter space is explored via changing $f^a$ to vary the P\'eclet number $Pe$ and $k_A$ to vary rigidity, the reversal frequency $\tau_n$, and packing fraction $\phi$.
\section{Results}
In this section we first discuss the impact of continuous change in the polar versus nematic driving on the unjamming of dense filaments ($\phi$=0.8). Next we study the impact of varying rigidity of the active elements on the unjamming process for both, polar and nematic driving. 
Finally, we show how different forms of activity allow the systems to achieve larger densities without getting jammed.

To quantitatively differentiate a jammed state from an unjammed state we employ the overlap function, $Q(t)$. If the system is in a jammed state, the displacements attained by individual filaments is very small. When the system undergoes unjamming transition filaments overcome the energy barriers and rearrange attaining larger displacements.
The overlap function is given by:
\begin{figure}[t!]
    \centering
    \includegraphics[scale=0.7]{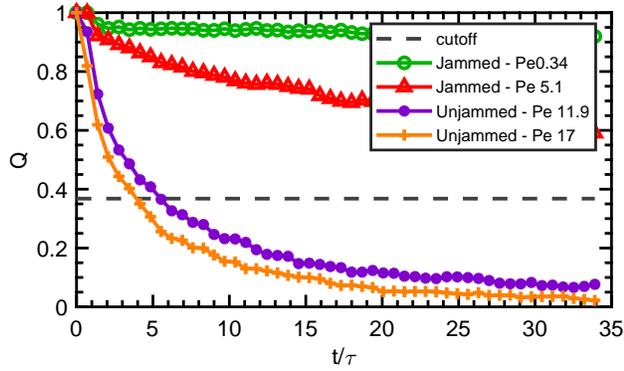}
    \caption{Overlap function $Q$ as a function of time $t/\tau$ for four different nematic activities, showing  active unjamming as the P\'eclet number $Pe$ increases. Rigidity is fixed at $k=48$. The horizontal dashed line shows the cutoff value $Q=1/e$.}
    \label{fig:Q}
\end{figure}
\begin{figure}[t!]
    \centering
    \includegraphics[scale=0.8]{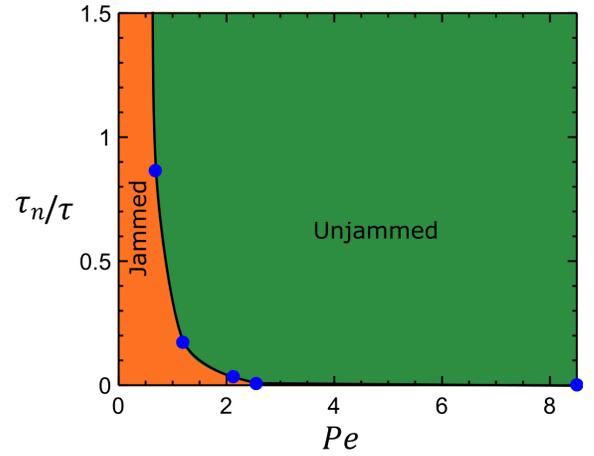}
    \caption{The phase diagram characterising the transition from the jammed to the unjammed state in the reversal time-activity, $\tau_n/\tau-Pe$, plane for the packing fraction $\phi=0.8$. At a fixed activity $Pe$ longer persistent time enhances the unjamming.}
    \label{tau_pd}
\end{figure}
\begin{equation}
Q(t) = \frac{1}{N }\sum_{i=1}^N w(|{\bf r}_i(t)-{\bf r}_i(0)|),
\label{Q}
\end{equation}
where, 
\begin{equation}
w(r) = \left \{  
\begin{array}{lr}
1,\quad r \leq d  \\ 
0, \quad r > d
\end{array}  \right.
\label{w}
\end{equation}
This function measures the fraction of filaments that have moved by more than a characteristic length $d$, which we choose to be half the length of a filament. A system is considered to be in the the jammed state if $Q(t)$ has not decayed below $\frac{1}{e} \approx 0.37$ at the end of the calculation period, where $e$ is the Euler's number. We have checked that the choice of this cutoff does not alter the results qualitatively. The overlap function is calculated for a period of approximately $35 \tau$. The system is given time to reach a steady state before the overlap function is calculated, ensuring the initial conditions do not have a large impact on the outcome. Examples of overlap function variation with time are demonstrated in Fig.~\ref{fig:Q} for four different nematic activities, showing active unjamming as the P\'eclet number $Pe$ increases. For $Pe=0.34$ (green open circles), and $5.1$ (red filled triangles) the system is in the jammed state, while for $Pe=11.9$ (violet filled circles), and $17$ (orange pluses) the system is unjammed as $Q$ decays below the cutoff value within the simulated time.
%
%
\subsection{The impact of the reversal time: polar and nematic drivings}\label{sec:tau}
\begin{figure}[t!]
    \centering
    \includegraphics[scale=0.6]{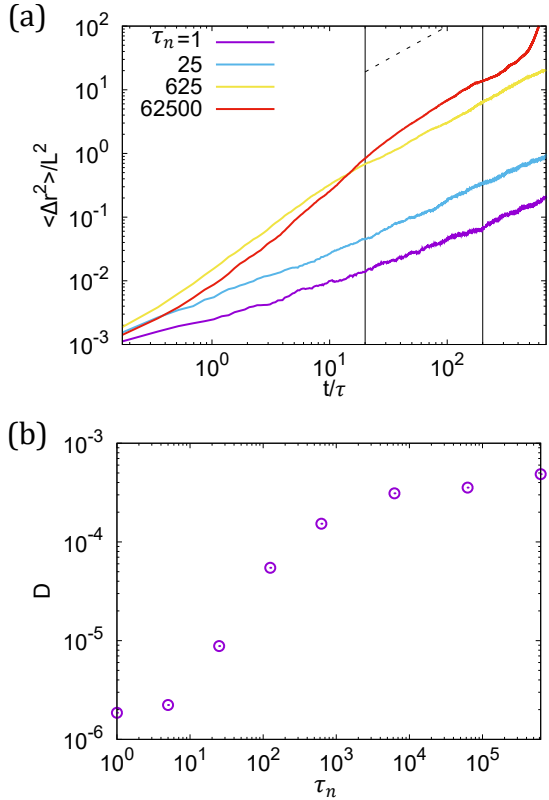}
    \caption{(a) Mean square displacement of a filament as a function of lag time for different reversal times and $Pe=0.85, \xi_p/L=48, \phi=0.8$. The grey-dashed line indicates slope of $1$. The vertical lines denote the range considered in finding the slope of MSD for the calculation of diffusion constant $D$, which is plotted as a function of the reversal time $\tau_n$ in (b).}
    \label{msd}
\end{figure}

We begin by exploring the impact of varying the parameter $\tau_n$, the reversal time period of active forces, on the transition from jammed to unjammed state. The two extremes of this parameter correspond to filaments that are either polar~\cite{duman_collective_2018} ($\tau_n > \tau_{\text{tot}}$) or nematic~\cite{joshi_interplay_2018} ($\tau_n = 1$), where $\tau_{\text{tot}}$ is the total simulation time. In this section, we keep the other parameters constant; the rigidity is fixed at $\xi_p/L = 48$ and the packing fraction at $\phi = 0.8$. Starting from nematic filaments, we incrementally increase the reversal time, until $\tau_n > \tau_{\text{tot}}$ that corresponds to the fully polar driving. As the reversal time increases, the system undergoes transition from a jammed to a unjammed states when all other parameters are fixed (Fig.~\ref{fig:1}). This is evident from the characterisation of jammed and unjammed state using overlap function $Q$ and measurements of mean square displacements (MSD) of the filaments (Fig.~\ref{msd}). 
Figure~\ref{tau_pd} presents a phase diagram of the transition activity for all the simulated reversal times, in the $\tau_n/\tau$-$Pe$ plane (Fig.~\ref{tau_pd}), demonstrating how the boundary between jammed-unjammed states is shifted towards lower activities as the reversal time is increased. This implies that for a high volume fraction $\phi=0.8$, at the same activity, polar driving force enhances unjamming compared to the nematic driving. However, as we will show later in section \ref{sec:density}, the efficacy of polar and nematic driving at unjamming the dense system changes when the system is at even higher packing fractions.
%


To further emphasise that increasing the reversal time or polar nature of the active driving enhances the unjamming of the system, we calculate the mean squared displacement $\langle \Delta r^2 \rangle$, for different reversal times at a fixed activity $Pe=0.85$. The scaled MSD as a function of lag time clearly shows a short-time sub-diffusive regime followed by a diffusive regime at late times for all the reversal times (Fig.~\ref{msd}a). We use the intermediate-time, as enclosed by the gray vertical lines in Fig.~\ref{msd}a, to calculate the diffusion constant $D=\Delta r^2/2t$ for all the curves and find that, $D$ increases with increasing reversal time (Fig.~\ref{msd}b). This provides a quantitative measure of the increased dynamics as the driving force becomes more polar in nature for the same filament activity. Interestingly, when plotted in a log-log plot the diffusion coefficient shows at least two distinct power-law scaling with varying the reversal rate: one at the intermediate reversal times (up to $\tau_n \sim 10^3$), followed by another slower increase for $\tau_n > 10^4$. This indicates that the effective temperature of the system $T_{\text{eff}}=\gamma D/k_{B}$ depends on the reversal rate for a fixed activity and can exhibit different dependence on the reversal time as the active forcing changes from nematic to polar driving.

In the next section we discuss the effect of varying filament rigidity on the jamming/unjamming phase transition.
\subsection{The impact of the filament flexibility on active unjamming}
\begin{figure}[b!]
    \centering
    \includegraphics[scale=0.5]{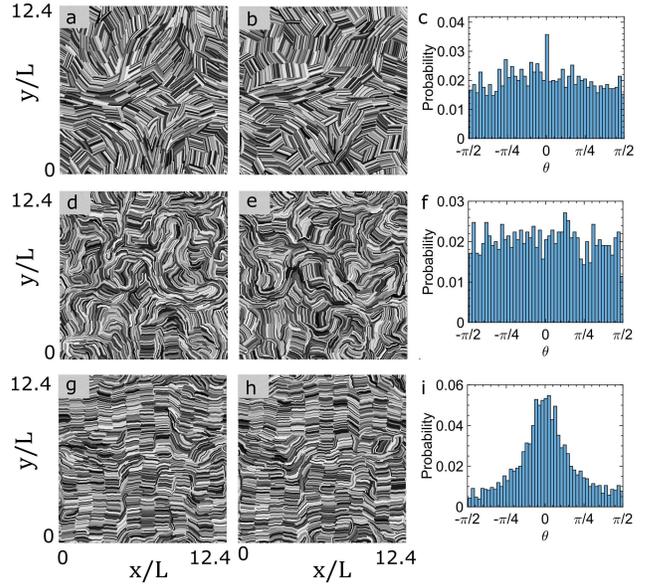}
    \caption{The re-entrant jamming behavior for polar active filaments. Snapshots for activity $Pe=0.34$ is shown for decreasing values of filament rigidity. Panels a and b show snapshots for $\xi_p/L = 768$ in the jammed state at times $t/\tau  = 173.0$ (a), and $ t/\tau = 276.8$ (b). Panels d, e show active turbulence for $\xi_p/L = 48$ at  times: $t/\tau = 173.0 $ (d) and $t/\tau  =  276.8 $ (e). In panels g, h snapshots for $\xi_p/L = 3$ in the jammed state are shown at times $t/\tau =2076.1$ (g) and $t/\tau= 2179.9.1$ (h). Panels c, f, and i plot the orientation distribution of filaments corresponding to snapshots in panels b, e and h, respectively. It is evident that there is strong alignment for $\xi_p/L =3$ (i) where as for higher values of $\xi_p/L$, the distribution is more uniform (c, f). The angle, $\theta$, is the relative angle between the filaments.}
    \label{kappa_pol}
\end{figure}
%
 Next, we study the impact of filaments flexibility, by varying the bending rigidity $\xi_p/L$ (see Eq.~\ref{Ubend}) of a filament on the jamming/unjamming transition. We vary $\xi_p/L$ between $\xi_p/L = 3$ to $\xi_p/L = 768$, with a higher value corresponding to a more rigid filament. As with the previous section, we keep the packing fraction fixed at $\phi=0.8$.  We do the same study for both nematic and polar filaments and find that the nature of jamming transition is very different in the two cases. For nematic driving filaments need monotonically higher activity to unjam as the rigidity is increased. In other words, higher rigidity promotes jamming for nematic filaments. However, this is not true for polar filaments. Rather, for more flexible filaments, i.e. $\xi_p/L<48$, we observe a re-entrant unjamming-jamming transition in the rigidity-activity phase space for polar driving. This can be seen in Fig.~\ref{kappa_pol}, which shows the polar system at activity $Pe=0.34$ transitioning from the jammed state at $\xi_p/L = 768$ (Fig.~\ref{kappa_pol}a,b), to unjammed state at $\xi_p/L = 48$ (Fig.~\ref{kappa_pol}d,e), to a re-entrant jammed state at $\xi_p/L = 3$(Fig.~\ref{kappa_pol}g,h). Here we show only two snapshots, one at $t/ \tau = 173.0$ (panels a, d) and the other at $t/ \tau = 276.8$ (panels b, e) for each $\xi_p/L$. In this way it is easier to observe the stark contrast between the jammed and the unjammed systems: while in panel (a), (b), (g), and (h), the snapshots look very much correlated and indicate  jammed states at very high and low rigidity, respectively, the snapshots in panel (d) and (e) look completely uncorrelated showing a unjammed state for intermediate rigidity. 

\begin{figure}[b!]
    \centering
    \includegraphics[scale=0.6]{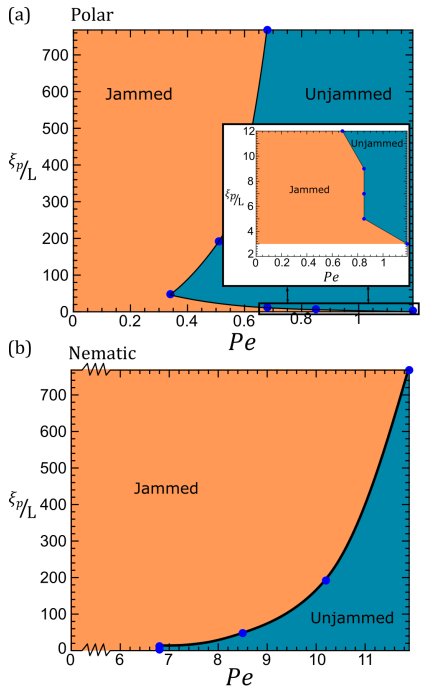}
    \caption{Rigidity-activity phase diagram for (a) polar and (b) nematic filaments with activity on the $x$-axis and rigidity on the $y$-axis. The inset in (a) shows a zoomed-in view at lower rigidity for polar filaments.}
    \label{kappa_pd}
\end{figure}
\begin{figure}[h!]
    \centering
    \includegraphics[scale=0.09]{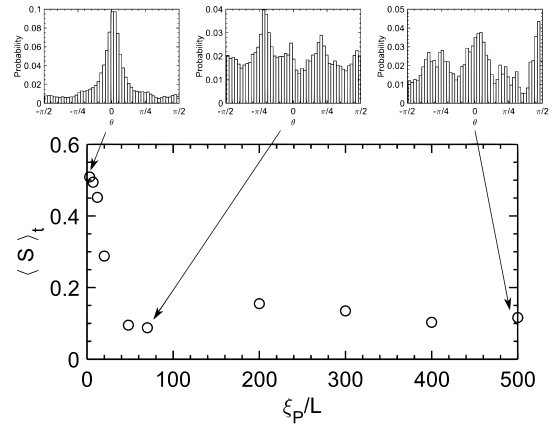}
    \caption{Time averaged nematic order parameter $\langle S \rangle_t$ as a function of rigidity $\xi_p/L$ at a fixed activity $Pe=0.4$. Three distributions of the coarse grained orientation $\theta_{\alpha}$ are shown for $\xi_p/L= 3$, $70$, and $500$.} 
    \label{reent_ord_prm}
\end{figure}

%
%
To show this re-entrant behavior more clearly, we construct phase diagrams for the polar (Fig.~\ref{kappa_pd}a) and nematic filaments (Fig.~\ref{kappa_pd}b) in the rigidity-activity phase space.  For polar filaments if the activity is very small ($Pe<0.34, \phi=0.8$) the system remains jammed irrespective of the filament stiffness. 
Interestingly, for larger activities ($\approx Pe>0.34$) we find a re-entrant transition as the rigidity is varied. Inset of Fig.~\ref{kappa_pd}a zooms in the phase diagram for $\xi_P/L<=12$. In this regime, as the rigidity is lowered, the system first undergoes an unjamming transition  at intermediate rigidity values followed by re-jamming at still lower rigidity. While the first transition is driven by softening due to reduced rigidity, the second transition is a cooperative effect of ordering, as evident from the snapshots and orientation distributions of filaments in Fig.~\ref{kappa_pol}. To show this emergent ordering at lower rigidity for polar driving, we define a nematic order parameter $S$:
\begin{equation}
    S = \left \langle \frac{1}{N_{cg}} \sqrt{ |\sum_{\alpha}^{N_{cg}} cos(2\theta_{\alpha})|^2 +|\sum_{\alpha}^{N_{cg}} sin(2\theta_{\alpha})|^2 } \right \rangle .
\end{equation}
Here the sum over $\alpha$ runs over $N_{cg}$ number of coarse grained boxes and $\theta_{\alpha}$ is the angle of the course grained director field of box $\alpha$. The brackets $\langle .. \rangle$ denote an average over multiple simulations. The value of $S$ grows initially before saturating at some steady state value depending on the rigidity. In Fig.~\ref{reent_ord_prm}, we plot this time-averaged saturated value, $\langle S \rangle_t$, as a function of filament rigidity $\xi_p/L$ at fixed activity $Pe=0.4$. As filaments become more rigid, the value of $\langle S \rangle_t$ rapidly falls until approximately $\xi_p/L <= 48$, after which it stays relatively constant. The dip at  $\xi_p/L = 48$ and $70$ can be attributed to the unjamming of the system at these values. Although $S$ can be used to categorise the transition at low rigidity, it remains unchanged at the jamming transition at higher rigidity.
%
\subsection{The impact of packing fraction on active unjamming}
\label{sec:density}
\begin{figure}[t!]
    \centering
    \includegraphics[scale=0.45]{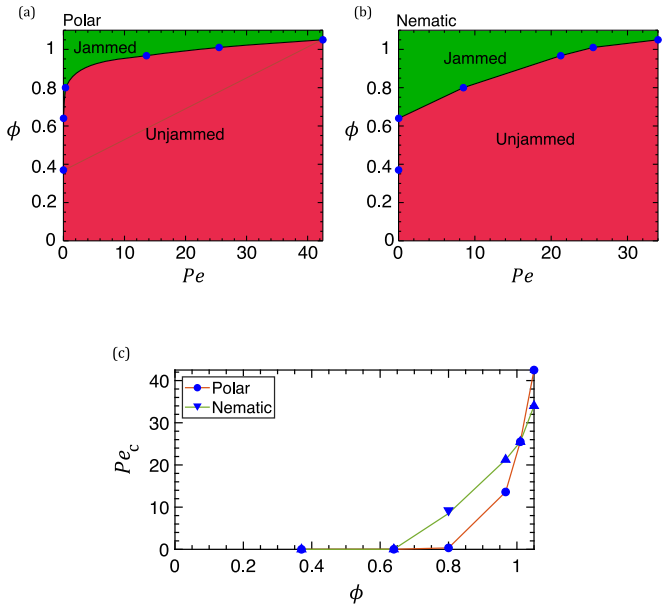}
    \caption{Phase diagram for both polar (a) and nematic (b) filaments with the activity on the $x$-axis and packing fraction $\phi$ on the $y$-axis. (c) Critical activity for the unjamming as a function of packing fraction for polar and nematic active driving.}
    \label{phi_pds}
\end{figure}

Having established the distinct impacts of polar and nematic driving on active unjamming and their subtle interplay with filament flexibility, we next explore the effect of varying the packing fraction $\phi$ of filaments on the dynamics, for polar and nematic driving at various activities, while keeping the filament rigidity fixed at $\xi_p/L = 48$. When $\phi$ is very small ($\phi < 0.64$) the system can unjam without any activity. The thermal noise is sufficient to drive the motion of the particles. For $\phi > 0.64$, the higher the density, a higher activity is needed to unjam the dense assembly of filaments. To investigate the impact of polar versus nematic driving at varying filament densities, Fig.~\ref{phi_pds} gives the two phase diagrams for polar (panel a) and nematic (panel b) filaments. It is apparent that a high activity is required to unjam both systems when $\phi$ is increased. For $\phi < 1.0$ nematic filaments require more activity to unjam compared to polar filaments as discussed earlier in sec.\ref{sec:tau}. Interestingly, however, this trend changes at higher densities $\phi \sim 1.0$, where we find that the activity required for unjamming of polar filaments becomes as much as that for nematic filaments and can even exceed it. This is best illustrated in Fig.~\ref{phi_pds}c, where we show the critical activity required for unjamming the system $Pe_c$, as a function of the packing fraction $\phi$ for polar and nematic systems. A crossover is evident close to $\phi = 1.0$ for the critical activity $Pe_c$ for polar and nematic filaments.\\
\begin{figure}[t!]
    \centering
    \includegraphics[width = 0.9\linewidth]{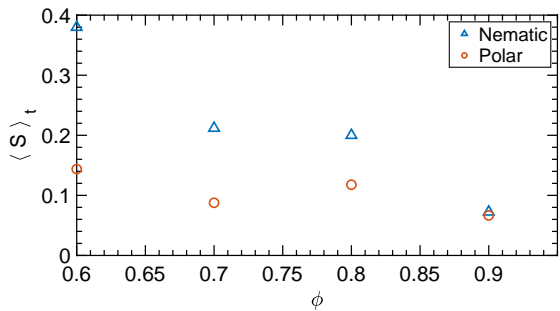}
    \caption{Time averaged nematic order parameter $\langle S \rangle_t$ as a function of packing fraction $\phi$ at a fixed rigidity $\xi_p/L=48$ and activity $Pe=0.34$ for polar and $Pe=6.8$ for nematic.}
    \label{nem_ord}
\end{figure}
We further calculate the time averaged nematic order parameter $\langle S \rangle_t$ as a function of packing fraction $\phi$ at a fixed rigidity $\xi_p/L=48$ and plot it in Fig.\ref{nem_ord}  for polar (red circles) and for nematic (blue triangles) driving. The plot clearly shows the dominance of nematic order for nematic driving at lower volume fractions. Moreover, it is evident that increasing the volume fraction suppresses nematic ordering in both polar and nematic driving cases.
\section{Discussion}
In this study we employed a generic discrete model for active flexible filaments to explore the interplay between different forms of activity with flexibility and its impact on jamming-unjamming transition in dense collectives of active matter. First, by varying the reversal rate of self-propulsion, we found that at a fixed activity, lower reversal rates are more effective at unjamming the system of active particles. These correspond to a more persistent directional self-propulsion and at the extreme of no reversal describe a purely polar self-propulsion, while the other extreme of fast reversals corresponds to the nematic (apolar) driving. Remarkably, at even higher densities we find that for a fixed activity, highly frequent reversal - nematic driving - becomes more effective than purely polar driving in unjamming the system. The crossover occurs close to the packing fraction $\phi \sim 1$ corresponding to the confluent state, where no inter-filament space exists between the active particles. The marked difference in the effectivity of polar and nematic driving in unjamming dense systems at different packing fractions can be explained based on the distinct modes of collective flux of filaments that polar and nematic driving generate: polar driving unjams the system by creating flocks of filaments that collectively move around the system~\cite{duman_collective_2018}, while nematic driving produces collective flows in the form of bend and splay patterns~\cite{joshi_interplay_2018}. For a fixed active forcing, at packing fractions below full confluency $\phi < 1$, the existence of even small inter-filament spaces allows for flocking patterns to emerge more easily relative to bend and splay deformations. On the other hand, at even higher densities when the system is fully confluent $\phi \sim 1$, polar flocks face more effective resistance from other, differently oriented, flocks in the system, while bend and splay deformations allow for filament flux to be generated relatively easier than the flocking-induced flux of particles. 

Such a density-dependent unjamming that crucially depends on the mechanism of self-propulsion can be relevant in biological setups, where switching between polar and nematic driving, e.g. by varying the reversal rate, can allow the collective to achieve active turbulent state even at very high densities. Since highly dynamic active turbulence state is more effective in enhancing the transport of mechanochemical signals or mixing of nutrients within the collective, it is possible that living organisms have evolved certain self-propulsion strategies that enables them to maintain the active turbulence at significantly higher densities compared to their inanimate counterparts. Similarly, under certain biological conditions, it might be favourable for the collective to be in a more solid-like, jammed, state and switching between polar and nematic self-propulsion modes can equip active materials to maintain such tunability.\\
Additionally, we found that when combined with the change in the self-propulsion mode, the flexibility of active elongated particles, can have a significant impact on the jamming-unjamming transition of the collective. In particular, for the polar driving at a fixed density and activity, while increasing the flexibility of rigid particles results in unjamming the system, a re-entrant behavior to a jammed state is revealed upon further increasing the flexibility. Interestingly, this second transition to jamming for highly flexible polar filaments coincides with the development of highly (orientationally) ordered domains within the system and the flexibility-induced emergence of nematic order. Similar to the jamming-unjamming transition that depends on the activity mode, this flexibility-dependent active transition can be relevant to biological setups and in particular in the competition between different phenotype of active cellular systems that could differ in their mechanical properties such as rigidity.\\
It is important to note that hydrodynamic interactions are not included in our model. We perform Langevin dynamic simulations which mimic the presence of implicit background solvent. Hydrodynamic interactions plays a crucial roles in many biological process. Recent study~\cite{martin-gomez_active_2019} of active polymers in a dilute regime shows that hydrodynamic interactions can significantly modify their conformational properties leading to substantial shrinkage of polymers and to shorter relaxation times due to enhanced mean square displacements. However, there are many biological systems such as neutral swimmers \cite{zottl_hydrodynamics_2014, gotze_mesoscale_2010, downton_simulation_2009}, swimmers near a wall \cite{yan_reconfiguring_2016, lowen_active_2018}, or microorganisms that glide on a surface \cite{di_leonardo_controlled_2016, nishiguchi_flagellar_2018}, for which hydrodynamic interactions are shown to be of negligible importance and hence our model can be particularly valid for. Moreover, most of our studies are performed at a very high packing fraction $\phi=0.8$, where the effect of hydrodynamic interactions is expected to be much less significant due to high filament concentration.\\
Finally, it is noteworthy that in this study we focused mainly on distinguishing the impacts of polar and nematic driving on the jamming-unjamming transition, without delving into the details of the collective flows, within the unjammed state, that are generated by the two distinct forms of activity. Previous work has demonstrated signatures of active turbulence in dense collection of polar active filaments by demonstrating the existence of power-law scaling of the kinetic energy spectrum with respect to the wave number~\cite{duman_collective_2018}. Future work should focus on the differences in the statistical imprints of the unjammed state in terms of possible flow modulations, e.g. power-law scaling of the energy spectrum and correlation length of vorticity field, as well as dynamics of topological defects within the unjammed state.

\begin{acknowledgments}
A. D. acknowledges funding from the Novo Nordisk Foundation (grant No. NNF18SA0035142 and NERD grant No. NNF21OC0068687, and Villum Fonden Grant no. 29476). C. M. acknowledges funding from Interaction fellowship and SERB (Ramanujan fellowship File No. RJF/2021/00012. We also acknowledge Dr. Abhijeet Joshi for helping in the implementation of direction reversal of filaments.).
\end{acknowledgments}

\section*{Data Availability Statement}
All the data and codes for analyses are available from the authors upon reasonable request.

\bibliography{bibfile.bib}

\end{document}